# *Exawatt-Zettawatt Pulse Generation and Applications*

G. A. Mourou[a], N. J. Fisch[b,*], V.M Malkin[b], Z. Toroker[b], E.A. Khazanov[c], A.M. Sergeev[c], T. Tajima[d]

[a] *Institut Lumière Extreme, ENSTA, Chemin de la Huniere, 91761 Palaiseau, France*
[b] *Princeton Plasma Physics Laboratory, Princeton University, Princeton, NJ 08543 USA,*
[c] *Institute of Applied Physics of Russian Academy of Science, 46 Uljanov St., N.Novgorod, 603950, Russia*
[d] *Faculty of Physics, Ludwig-Maximillians-Universität München, 85748, Germany*

*Abstract*: A new amplification method, weaving the three basic compression techniques, Chirped Pulse Amplification (CPA), Optical Parametric Chirped Pulse Amplification (OPCPA) and Plasma Compression by Backward Raman Amplification (BRA) in plasma, is proposed. It is called $C^3$ for Cascaded Conversion Compression. It has the capability to compress with good efficiency kilojoule to megajoule, nanosecond laser pulses into femtosecond pulses, to produce exawatt and beyond peak power. In the future, $C^3$ could be used at large-scale facilities such as the National Ignition Facility (NIF) or the Laser Megajoule (LMJ) and open the way to zettawatt level pulses. The beam will be focused to a wavelength spot size with a f#1. The very small beam size, i.e. few centimeters, along with the low laser repetition rate laser system will make possible the use of inexpensive, precision, disposable optics. The resulting intensity will approach the Schwinger value, thus opening up new possibilities in fundamental physics

Keywords: ultrafast lasers, high energy laser pulses, relativistic optics, plasma optics

## 1. Introduction

Last year the 50th anniversary of the laser was celebrated. The laser, by its coherence, monochromaticity and field magnitude, was the gateway to novel spectroscopic methods of investigation that deepened our understanding of the atomic structure. However, the laser was inefficient to probe the subsequent strata formed by the nucleus, the nucleon or the vacuum. Neither the laser photon energy nor its electric field has been large enough to conceive decisive experiments beyond the atomic level.

To reach the level where relevant nuclear and/or high energy physics investigations could be undertaken, a new type of large-scale laser infrastructure was conceived under the aegis of the European scientific community. The goal of the new infrastructure named Extreme Light Infrastructure (ELI) [1] is to yield the highest peak power and laser focused intensity. With its peak power of 200 PW, ELI represents the largest civilian laser project in the world. This gargantuan power will be obtained by releasing few kJ in 10 fs. Focusing this power over a micrometer spot size will yield intensities in the $10^{25}$ W/cm$^2$ range, well into the ultrarelativistic regime (see figure 1). This extremely high peak power will lead to the highest electric field, but also according to the pulse intensity-duration conjecture [2] to the shortest pulse of high-energy particles and radiations, in the attosecond-zeptosecond regime. The particle energy could be in the 1-1000 GeV regime, enough to give access to Nuclear Physics, High Energy Physics or Vacuum Physics. ELI will enable to reach the entry point into this new regime. To get deeper as shown in Fig. 1, will demand intensities two or may be three orders of magnitude higher than what is planned for ELI.

Over the last 25 years our ability to amplify laser peak power has been remarkable and has revolutionized laser science. Three amplification techniques have been demonstrated. The first in 1985 was CPA [3]. Today, it is the workhorse of most laser amplification systems. The second has been the OPCPA [4]. Similar to CPA in concept, it relies on the amplification of light by Optical Parametric Amplification. It is used for broadband, few cycles pulse amplification. Finally, about 15 years ago, a new compression technique based on Backwards Raman Scattering (BRA) was presented [5]. It offers the distinct advantage of avoiding gratings, the major hurdle in ultrahigh high peak power pulse amplification.

The regime beyond the exawatt, towards the zettawatt is the next high intensity frontier. As pointed out as early as 2002 [6], it could be realized by compressing the energy delivered by NIF or LMJ-like laser, from ns to 10 fs duration pulses. However, the scheme proposed to produce and compress MJ pulses was based on CPA and OPCPA alone, and thus unfortunately uses unrealistically large and expensive diffraction gratings. A two-stage method, first using CPA (or OCPA) for compressing a pump beam from ns to ps and then in a second stage using BRA to fs compress the ps output was described in 2003 [7]. In 2005 [8], an alternative two-stage method of compression from ns to fs was described, where each stage is a BRA stage; however the first BRA stage employs a variation of the BRA scheme wherein the BRA occurs at an ionization front [9, 10].

None of the existing compression techniques alone is really suited to the amplification and compression of 10 kJ to MJ pulses. In this letter, we specify a combined approach which uses the three basic techniques to make possible efficient compression of 10 kJ to MJ, 5 ns pulses into fs pulses of the highest quality. This technique employs CPA to produce a first-stage compression, followed by BRA for a second-stage compression, together with OPCPA to produce a strong seed pulse. The combination of the three is optimized over the joint compression technologies of both compression stages. We call this optimized approach $C^3$, for Cascaded Conversion Compression. In order to provide the highest intensity a novel f#1 focusing approach is described.

## 2. Beyond the Exawatt Concept

To reach the exa- to zettawatt level, we need to produce at least of the order of 10 kJ in 10 fs pulses. As mentioned earlier, none of the existing compression techniques mentioned can do that individually. For instance, if we start from a 10 kJ level, 5 ns pump pulse and want to compress it to 10 fs by CPA or OPCPA, there will be a serious problem with the diffraction gratings. For this application, we need to use a broadband (200 nm) metal-coated gratings. These gratings [11] have a low damage threshold on the order of 100 mJ/cm$^2$. A basic CPA/OPCPA arrangement will therefore require a costly grating area as large as 10 to 100m$^2$, formed by tiling of the order of 100 gratings of one m$^2$ area. If we use the Plasma Compression Technique alone, BRA, it will require also a plasma with an unrealistic length given by half the pulse duration times the $c$, the speed of light.

The rule of thumb here for Backward Raman Amplification imposes that the plasma must be uniform over a length given by half the laser pulse length. This corresponds to plasma 75 cm long for a 5 ns pump pulse, with several centimeter aperture. Getting good plasma uniformity over this entire volume is obviously a tall order.

The philosophy behind $C^3$ is therefore to break down the amplification technique in three steps Involving the CPA, OPCPA and compression by BRA. To explain the concept we will use as an example the laser LIL, at the CEA (France), composed of 8 beams, each one delivering energy around 10 kJ in 5 ns at 1.05 µm. To these pulses we will add a slight frequency chirp corresponding to $\Delta\omega = 1/20$ ps which will make the compression of the pulse possible to around 20 ps.

## 2.1 First Step: CPA Compression

The first step utilizes the CPA technique. Working on one beam, we compress the 10 kJ, 5 ns, chirped pulse into a 10 kJ, 20 ps, Fourier-transformed pulse. Because the desired final pulse duration from the CPA stage should be about 20 ps, corresponding to a narrow bandwidth, we can conveniently use dielectric gratings instead of metallic-coated ones. Dielectric gratings are important for two reasons. First, for narrow bandwidth, they have an excellent efficiency of 99%. Second, because the pulse is rather long, i.e. 20 ps, the gratings have a much higher damage threshold than metal-coated ones. The damage threshold for dielectric for 20 ps is around 5J/cm$^2$ while for metal-coated ones it is 0.1 J/cm$^2$ for 10-20 fs pulses. We recall that for the regime ns to 1ps the damage threshold in dielectric varies [11] as the $T^{1/2}$, where T is the incident pulse duration. This represents an enormous improvement of 50 times, which translates linearly in total area and cost. Third, they are commercially available in a square-meter size. As we will see later, this intermediate step compression will take only of the order of 0.2 m$^2$ area gratings for 10 kJ.

## 2.2 Second Step: Generation of an Intense and Clean Seed Pulse by OPCPA

The second step requires us to generate a strong femtosecond seed pulse at 1250 nm, in order to facilitate the BRA process in step 3, and for which OPCPA is superbly adapted. This point is important because during the amplification process the BRA signal produced by the interference between the pump and the seed signal, will compete with several processes that arise from noise: the fast Raman Back Scattering (RBS) caused by the pump scattering on the thermal Langmuir waves on its way to the seed pulse [12]; the formation of precursors that may lead to premature pump depletion [13]; the scattering off density inhomogeneities [14]; and the resonant side-scattering process that arises under large cross-section plasma [15]. These deleterious processes can be thwarted by detuning the resonance via the introduction of suitable density gradients with suitable chirping of the pump (for example as described in [12]). Interestingly, multiple pumps can also be used, and suitable detuning can be introduced through adjusting the multiple pumps [16]. In each of these cases, as the unwanted interactions are suppressed, a strong seed will allow the resonant desired interaction to dominate through nonlinear processes. The reason that this selectivity is possible is because the unwanted interactions, arising from noise or otherwise in a linear regime, are more sensitive to detuning than are the desired nonlinear interactions.

OPCPA is eminently suitable to obtain the largest possible input seed pulse ($I_s$). A. Sergeev and his group have already demonstrated the generation of laser pulses of about

50 J, about 20 fs duration, and tunable in the 1250 nm regime by using the radiation from a Cr:Forsterite master oscillator as the seed for the OPCPA [17]. As shown, 1250 nm is the idler wave and in order to satisfy ultra-broadband phase matching an angular chirp must be induced in the beam by means of a prism placed between stretcher and OPCPA. After the OPCPA the idler wave (on the contrary to signal wave at 910 nm) keeps this angular chirp, which will be canceled by special design of a compressor.

As in any OPCPA laser a precise synchronization of pump and chirped pulses is required. It may be done either electronically or optically. Here, we will use a small fraction, i.e. 0.5 kJ of the compressed signal after frequency doubling to produce by Optical Parametric Generation a pulse around 10 fs, 100 J and 1250 nm. The latter will be used as the seed pulse in the next step.

**2.3 Third step: Compression by Light Conversion through Backward Raman Scattering**

In the third step, we will use the Backward Raman Amplification (BRA) method to further reduce the pulse duration from ~20 ps to ~10 fs. The Raman amplification amplifies radiations at frequency $\omega_s = \omega_L - \omega_p$, where $\omega_s$ is the Raman frequency, $\omega_L$ the laser frequency and $\omega_p$ the plasma frequency given by $\omega_p = \left(4\pi e^2 n_0 / m\right)^{1/2}$. Here $e$ and $m$ are the charge and mass of the electron and $n_0$ is the plasma electronic density. For our example, the pump wavelength is at 1050 nm. For a plasma frequency of $n_0 = 10^{19}$ cm$^{-3}$ the Raman wavelength will correspond to 1250 nm. The pump pulse (10 kJ, 20 ps, 1050 nm) and the seed (signal) pulse of ~10-20 fs duration at 1250 nm collides in a L ~ 3 mm long plasma cell corresponding to $T_p$ x $c/2$. $T_p$ is the pump pulse duration and c the speed of light in the plasma. The pulse intensity $I_P$ must be of the order of $10^{17}$ W/cm$^2$, corresponding to a fluence $F_p$ ~ 1 kJ/cm$^2$. It imposes a cell diameter of about 3 cm for a 10 kJ beam. It is worth noting that the plasma compression technique is less insensitive to the number and the direction of the pump beams than the optical methods utilizing crystals, as the plasma does not have its intrinsic axes of symmetry like in crystals [17]. However, as the Raman process operates close to the wavebreaking limit, we may have unwanted nonlinear effects, or we may have to compromise the efficiency to avoid the unwanted nonlinearities. Therefore it could be possible to pump the plasma cell with the 8 LIL beams, coming at different angles. The volume grating produced by the interferences between the signal and the pump will diffract the light in the signal direction (Fig. 3). It represents a simple way to multiplex and compress the 8 beams into one beam of 10-20 fs and 100 kJ. Conceptually this scheme could be extended to the NIF and the Laser Megajoule with 200 beams.

**3. Optimization from a Three Case study**

To illustrate the C$^3$ concept we present three cases involving a single pump beam, where a plasma slab mediates the Raman compression. The pump pulse is obtained by compressing the 5 ns pulse using CPA into either a 40 ps, 20 ps, 11 ps or 2 ps pulse. The OPCPA will provide a seed pulse of 20 fs, and 11 fs. Three plasma densities, $10^{19}$, $2\times10^{19}$ and $10^{20}$ /cm$^3$ have been studied. At these plasma densities, Landau damping and inverse bremsstrahlung are considered as is. The self-consistent heating of the plasma, dispersion effects of the waves are also taken into account. The plasma is assumed to be carbon, which in mediating the wave interactions becomes fully-ionized.

### 3.1 Case One

The plasma density is $10^{19}/cm^3$. For a seed pulse of 20 fs duration, this case is optimized by a 40 ps pump with intensity of $1.2 \times 10^{14}$ W/cm$^2$. The plasma length is 6 mm. The output pulse is 50 fs with 50% efficiency. The pump fluence is 4.8 kJ/cm$^2$, meaning that plasma with a cross-section of 2 cm$^2$ will process 10 kJ. A 20 fs input seed produces a 50 fs output pulse of 50 PW/cm$^2$ or a fluence of 2.5 kJ/cm$^2$. Thus the efficiency of energy conversion is about 50%. Clearly a shorter input pulse is advantageous in compressing to shorter output pulse, but it will require higher density plasma. This will be explored in the two following cases.

### 3.2 Case Two

Mediation is done by a plasma with a density of $2 \times 10^{19}/cm^3$. This case is optimized by a 20 ps pump pulse with intensity of $2 \times 10^{14}$ W/cm$^2$. The plasma length is 3 mm and the output pulse is 35 fs with 46% efficiency. However, with a shorter input pump duration, the output can be somewhat shortened. For pump intensity $4.5 \times 10^{14}$ W/cm$^2$, and an input pulse of 11 ps, the output pulse will have a duration of 26 fs with a 24% efficiency.

### 3.3 Case Three

Mediation by a plasma density of $10^{20}/cm^3$. This case is optimized by a pump pulse of 2 ps with intensity of $4.6 \times 10^{14}$ W/cm$^2$. The plasma length is about 6 mm. The output pulse duration is 20 fs with an efficiency of 24%. Keeping the same density shorter output pulses can be achieved. For an 11 fs input pulse we can produce a 20 fs output pulse. The pump requirement is 2 ps, so that the plasma length is about 0.3 mm. Here, the input intensity is 2 PW/cm$^2$ and relatively independent of the input seed pulse from 1 to 100 TW/cm$^2$. The output pulse amplified seed is about 46 PW/cm$^2$, corresponding to an efficiency of 24%.

### 3.4 Results

Table 1 summarizes the results of the three case studies. These results lead to the remarkable conclusion that it is indeed possible to compress 10 kJ pulse by using the C$^3$ technique to 10 to 20 fs with good efficiency (50%). The three techniques combined provide an overall compression factor of almost $10^6$. The grating size for the CPA part—reasonable and affordable—will be less that 1 m$^2$ for a damage threshold of 5 J/cm$^2$. The plasma size would also be of the order of 2 cm$^2$. The technique has also the capability to compress the 8 different beams coming from different directions into a single pulse of about 100 kJ (Fig. 3). This step opens the way to pulse compression at the MJ level provided by NIF or the Laser Megajoule.

### 4. Laser Focusing to Single Wavelength Spot Size

The C$^3$ technique makes possible the generation of high energy density beams of the order of few kJ/cm$^2$ as opposed to J/cm$^2$ in common cases. To attain the highest intensity we will need to focus the beam to a wavelength limited diffraction spot of the order of micrometers for a f#1 optics. This is a very tall order and necessitates extremely large f#1reflective optics. The surface area and diameter would be respectively $10^5$ cm$^2$ and > 1 m in diameter to accommodate 10 kJ for a ~1J damage threshold. If we are using the embodiment proposed in [7] considering the ratio, plasma thickness of ~1mm, over the aperture of few centimeters, i.e. 2-5, only spherical waves corresponding to f# 20-50 could be amplified, leading to a focal spot size of 10-50 wavelengths (or micrometers) producing a drop in intensity of 1000 times over what could be expected for f#1 optics. This effect could totally thwart the power gain obtain by temporal compression.

We propose, to attain the highest obtainable intensity, i.e. $10^{26}$ W/cm$^2$ for an exawatt pulse, by using, precision, disposable, centimeter size f#1 optics. The fluence on the mirror of 1-3 cm diameter could be of the order of 1-5 kJ/cm$^2$ for a 10 kJ. The mirror could be bear, gold or Al-coated. It will be exposed over the all area to an intensity of the order of $10^{18}$ W/cm$^2$ producing a highly reflective plasma. As mentioned previously it is vital to produce a high contrast pulse, so no pre-plasma is formed before the main pulse arrival on the mirror. Small size high precision optics could be mass-produced and replaced after each shot. The time between shot could be from 10 to 60 minutes. An inexpensive Taylor-made phase plate could also be introduced into the beam to perfect the wavefront correction. As with the mirror, they would need to be changed after each shot.

## 5. The C$^3$ a novel paradigm in Fundamental Physics

The C$^3$ technique applied to MJ systems may enable the exploration of fundamental physics at a depth that far surpasses what we see today [2]. For example, the current laser wakefield acceleration (LWFA) based collider concept is designed for a plasma density around $10^{17}$/cm$^3$, with a laser of 30-100 J for a stage, requiring 40 MW total laser power for the eventual collider [18]. If we can adopt the plasma at the lower density of $10^{15}$/cm$^3$ for example, the necessary average laser power, the toughest requirement, in a collider would dramatically decrease due to the 3/2 power scaling of the plasma density [19]. On the other hand, as the laser energy per stage scales inversely proportionally with the 3/2 power of the density, this would require much higher energy laser (10 kJ per stage). Having no such laser available currently, perhaps the adoption of higher density design would be understandable. However, there is not only the laser power requirement but also other beam handling requirements, such as the emittance control. These are more relaxed for the low-density / high energy laser operation of LWFA if the current technique can provide large energy, low power lasers.

As yet another example of a fundamental physics application, we consider the probing of a possible nonlinearity of vacuum due to fundamental new light-mass weak coupling fields such as dark matter and dark energy. With a 100 kJ class laser we foresee the search power to be as sensitive as even the level of the weakest gravitational coupling [20]. This is because our search capability (i.e. the signal) for the new fields increases as a function of the laser energy cubed and inversely proportional to the pulse length of the laser. As exemplified by these two prominent fundamental applications, the advancement of intense short-pulsed laser energy by 2-3 orders of magnitude empowers us a tremendous potential of unprecedented discoveries. These include: TeV physics, new light-mass weak-coupling field discovery potential, nonlinear QED and QCD fields, radiation physics in the vicinity of the Schwinger field, and zeptosecond dynamical spectroscopy of vacuum.

**Acknowledgements**

The work was supported, in part, through DOE Research Grant No. DE-AC02-09CH11466 and through the NNSA SSAA Program through DOE Research Grant No. DE274-FG52-08NA28553, Gérard Mourou was supported by the Fondation de l'Ecole Polytechnique and ELI 212105 T.T. was supported in part by the Blaise Pascal Foundation and by Deutsche Forschungsgemeinschaft Cluster of Excellence MAP (Munich Centre for Advanced Photonics). The authors wants to thank Natalia Naumova and Bianca Jackson for helpful discussions.

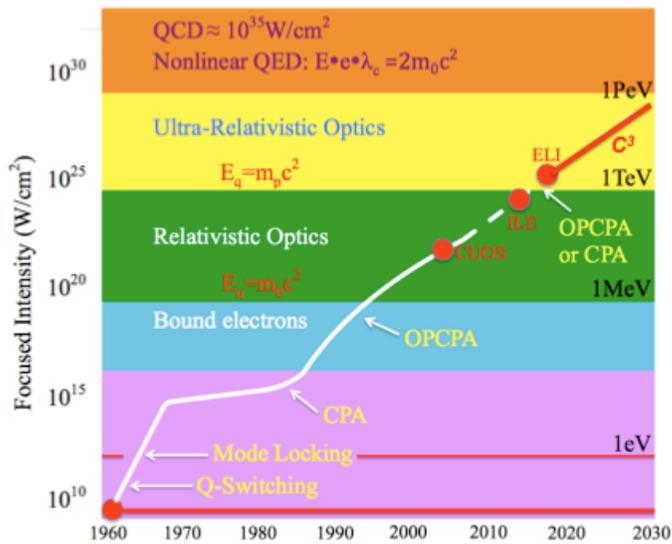

Fig.1 Graph showing laser focused intensity vs years. It is shown the highest peak power laser planned today, namely ILE and ELI. They are typically 10 PW and 100 PW facilities providing access to the ultrarelativistic interaction regime. The $C^3$ in combination with large scale pump laser would make possible the generation of exawatt and possibly zettawatt pulses extending laser matter interaction to vacuum nonlinear and high energy particle physics.

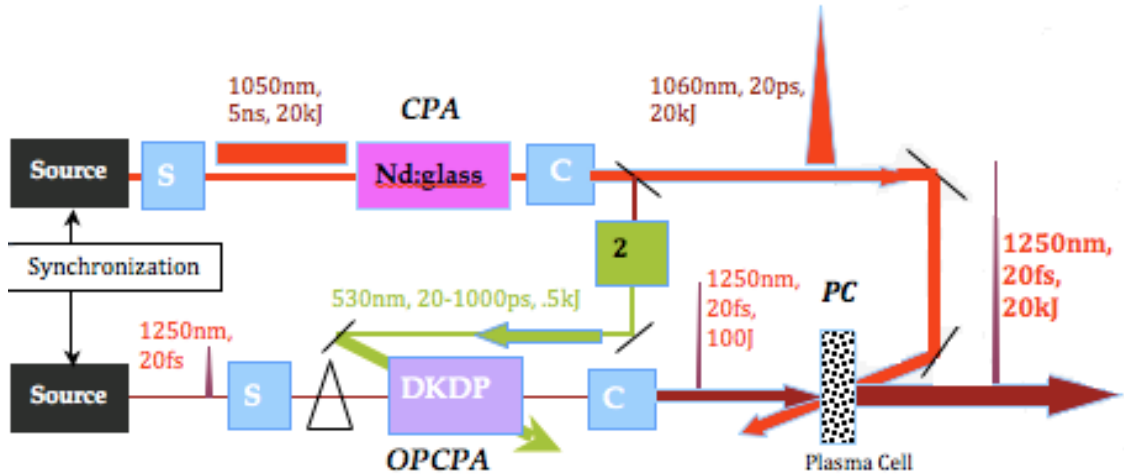

Fig. 2: Diagram showing the compression technique C³ resulting from the cascaded actions of the three basic techniques, CPA, OPCPA and PC. The CPA technique compresses 5ns, 20kJ into a ~20ps pulse. This pulse is used after frequency doubling, to pump an OPCPA. A strong idler wave is produced at 1250nm. The latter seeds, the plasma compression cell where by interfering with the 20ps pump pulse at 1050nm converts and transfers the pump into the seed pulse at 1250 nm. To preserve the pulse shortness, a prism in the OPCPA is used to produce the necessary angular chirp.

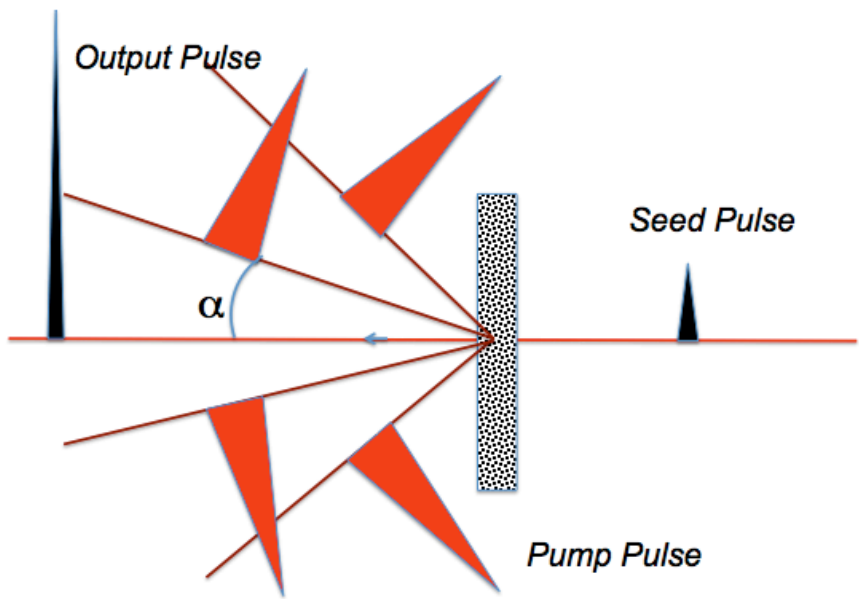

Fig. 3. Diagram displaying the concept of Multiple-Beam-Pumping showing that the energy from several beams can be transferred to one signal seed pulse.

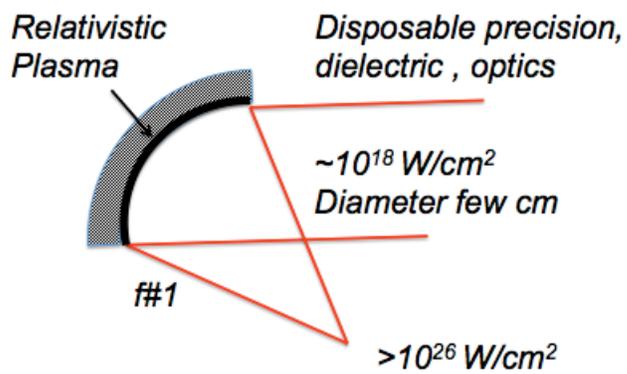

Fig.4. Diagram depicting the focusing of extreme light using a plasma disposable mirror.

| Case | $n_0$ (1/cm³) | $T_{s,in}$ (fs) | $T_p$ (ps) | $I_{p,in}$ (PW/cm²), $F_{p,in}$ (kJ/cm²) | Plasma length (mm) | $T_{out}$ (fs) | $I_{out}$ (PW/cm²), $F_{out}$ (kJ/cm²) | Efficiency (%) |
|---|---|---|---|---|---|---|---|---|
| 1 | $10^{19}$ | 20 | 40 | 120 | 6 | 50 | 2.4 | 50 |
|   | $10^{19}$ | 20 |    | 4.8 |   | 50 | 50 |   |
| 2 | $2 \times 10^{19}$ | 20 | 20 | 200 | 3 | 35 | 50, 2 | 46 |
|   |   | 20 | 11 | 450, 5 |   | 26 | 84, 2.2 | 44 44 |
| 3 | $10^{20}$ |   | 2 | 460, .9 | 6 | 20 | 46 | 24 |
|   |   | 11 | 2 | 2000, 4 | .3 | 20 | .92 | 24 |

Table 1: Summary of the results of the three case studies.